\documentclass[12pt]{article}
\usepackage{amsmath,amssymb}
\usepackage{scicite}
\usepackage{times}
\usepackage{hyperref}
\usepackage{color,gensymb,amsmath,amssymb,graphicx,morefloats}
\usepackage{bm}
\usepackage{graphicx,morefloats}
\usepackage[resetlabels]{multibib}
\usepackage{subfigure} 
\usepackage{color,soul}
\usepackage{caption}

\newcites{SM}{Supplementary References}

\newcommand{\kk}{{\bm{k}}}

\newcommand{\RR}{{\bm{R}}}
\newcommand{\rr}{{\bm{r}}}

\newcommand{\be}[0]{\begin{equation}}
\newcommand{\ee}[0]{\end{equation}}
\newcommand{\ba}[0]{\begin{eqnarray}}
\newcommand{\ea}[0]{\end{eqnarray}}

\newcommand{\up}[0]{\uparrow}
\newcommand{\dn}[0]{\downarrow}
\newcommand{\bmat}[0]{\begin{bmatrix}}
\newcommand{\emat}[0]{\end{bmatrix}}

\begin{document}

\thispagestyle{empty}

\hyphenation{va-ni-sh-ing}



\baselineskip24pt

\begin{center}
\vspace{-1.5cm}

{\Large 
Coupled topological flat and wide bands: Quasiparticle formation and destruction
}
\\[0.6cm]

\normalsize{Haoyu Hu
and Qimiao Si$^{\ast}$}
\\
[0.1cm]

\small\it{
Department of Physics and Astronomy, Rice Center for Quantum Materials, Rice University,
Houston, Texas, 77005, USA
\\[0.1cm]
}

\end{center}

Flat bands amplify correlation effects and are of extensive current interest.
They provide a platform to explore both topology in correlated 
settings
and correlation physics enriched by topology.
Recent experiments in correlated kagome metals 
have found evidence for
strange-metal behavior.  
A major theoretical challenge is to study the effect of local
Coulomb repulsion when the 
band
topology obstructs a real-space description.
In a variant to the kagome lattice,
we identify 
 an orbital-selective Mott transition
for the first time in any system
of coupled topological flat and wide bands.
This was made possible by the construction
of exponentially localized and Kramers-doublet Wannier functions,
which in turn leads to an effective Kondo lattice description.
Our findings show how quasiparticles are formed in 
such coupled topological flat-wide band systems and,
equally important, how they are destroyed.
Our work provides a conceptual framework 
for the understanding of the existing and emerging 
strange-metal properties
in kagome metals and beyond.



\newpage

\noindent{ {\bf\large Introduction}} \\ 
In flat electronic bands, Coulomb interactions are proportionally enhanced 
due to their reduced kinetic energy. As such, flat-band systems serve as 
a platform for strong correlation physics \cite{Pas21.1}.
Heavy fermion metals represent a canonical case of flat bands formed from 
highly localized atomic orbitals, and indeed they display rich 
correlation physics such as quantum criticality, strange metallicity and unconventional 
superconductivity \cite{Coleman-Nature,Kirchner2020}.
Studies here have led to the notion that 
quasiparticles are lost in strange metals 
\cite{Hu-qcm2022,Phillips-science22,Si2001,Colemanetal,senthil2004a}.
Another case, emerging 
in a growing list of materials, corresponds to flat bands formed 
by geometrical interference \cite{Mielke1991,Checkelsky2018,Yao18.1x,Comin2020-2,
flat_band_kagome_5
};
such bands are often topological.
These materials represent a playground to study
the both the strong correlations and topology~ {\cite{Setty2021,Setty2022,flat_band_kagome_2,flat_band_kagome_1,flat_band_kagome_3}}
, and have been
found to show unusual properties such as exotic forms of charge-density-wave order
\cite{Hasan2020,Guguchia2022-TRSB,Zhou2021.2,Dai2022,Yin2022,Setty2022}.
Recently, experimental evidence for strange-metal behavior has emerged 
from correlated kagome metals \cite{Ye2021.x,Ekahana2021.x},
which contains both flat and wide bands that intersect with each other.
The observed behavior takes the form
of a $T$-linear resistivity or a single-particle damping rate that is linear in frequency.
This raises the question of how quasiparticles can be destroyed in such systems.
The question is not only important for correlated kagome metals, but also for related flat band settings such as moir\'e systems \cite{Jao22,Zhao2022.2}.

Thus motivated, here we study 
the 
topological flat bands coupled 
to wide bands. Our work provides the first 
theoretical demonstration of an orbital-selective 
Mott transition in any system that involves 
coupled topological flat-wide bands. 
The orbital-selective Mott transition provides 
a framework to understand the
quasiparticles' formation 
and, equally important,
their
destruction.
The latter 
allows for the understanding of 
the existing and emerging 
strange-metal properties of flat-band systems.
Our work also connects the topological flat band systems to 
the orbital-selective physics of
bulk correlated materials \cite{Huang2022,Yu2021,Hu2022.2}, in which the atomic orbitals exhibit different dispersions and yet [in contrast to
 models with fully-decoupled orbitals \cite{Anisimov2002}]
are kinetically coupled 
with each other.
 
More specifically,
the topological nature of the flat bands 
makes it difficult to describe them
in terms of any localized orbitals.
The band topology obstructs the 
formulation of exponentially localized and Kramers-doublet Wannier orbitals  \cite{Vanderbilt2012-RMP}.
This poses a challenge for treating
 the effect of sizable spatially-local Coulomb repulsion,
as well as for connecting
the behavior of the coupled topological-flat-wide bands to the orbital-selective Mott physics.

\begin{figure}[t!]
    \centering
    \includegraphics[width=1.0\textwidth]{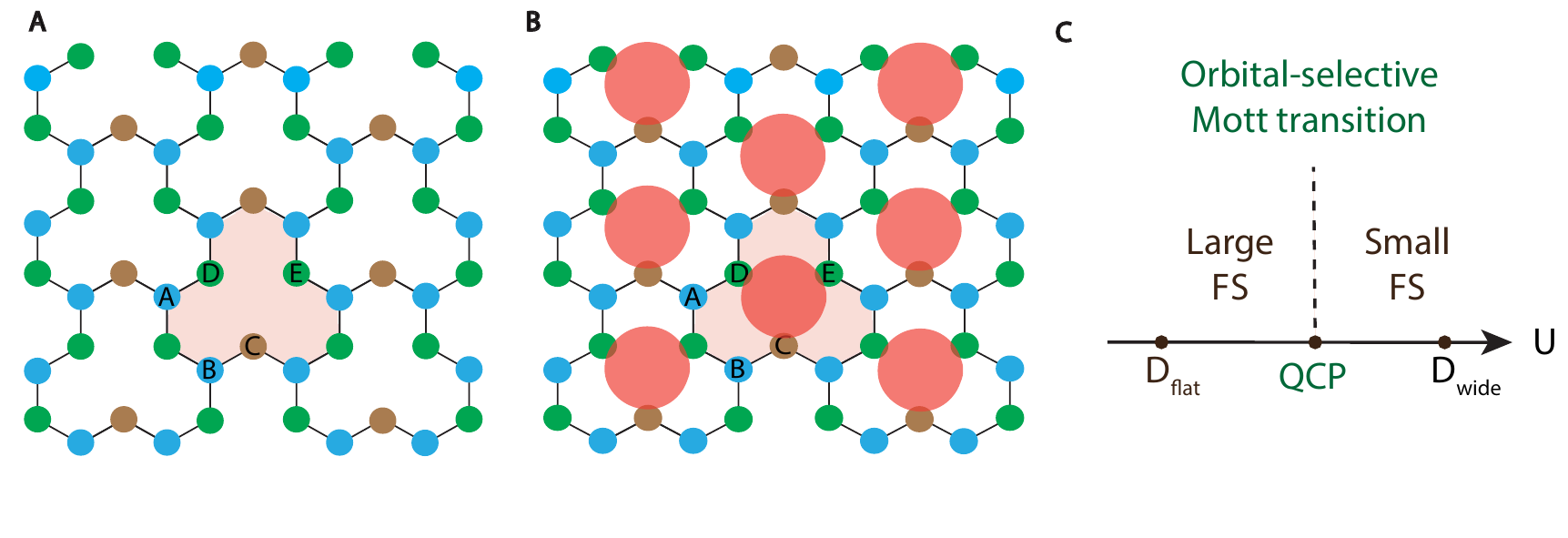}
    \caption{
    {\bf Lattice geometry and qualitative phase diagram.}
     ({\bf A}) Lattice geometry. 
     $A,B,C,D,E$ mark the five
    sites per unit cell.
    ({\bf B}) The Wannier orbitals 
    we construct, which form a triangular lattice 
   (the orange dots).
  ({\bf C}) Illustration of the zero-temperature phase diagram that we determine,
  for the Hubbard interaction that is larger than the width of the flat band ($D_{\rm flat}$) and smaller than the width of the wide bands ($D_{\rm wide}$). 
 }
 \vskip -0.5cm
    \label{fig:disp}
\end{figure}

Here, we make progress by considering a
lattice 
with a lower symmetry, which
retains the central issue of topological obstruction 
while allowing -- in a particularly transparent way -- for the construction of 
exponentially localized and Kramers-doublet Wannier orbitals.
The lattice, as illustrated in Fig.\,\ref{fig:disp}A, is a variant of the kagome lattice.
We are able to construct the Wannier centers, which 
turn out to form a triangular lattice (Fig.\,\ref{fig:disp}B).
When the effective local Coulomb repulsion is larger than the width 
of the flat 
band ($D_{\rm flat}$) but smaller than that of the wide band ($D_{\rm wide}$), 
we identify a continuous orbital-selective Mott transition 
[a quantum critical point (QCP)] (Fig.\,1C).
The two involved ground states respectively feature a 
``large" 
and a ``small" Fermi surface;
in an effective Kondo-lattice formulation that we construct,
the large Fermi surface is Kondo driven;
the small Fermi surface, then, develops from a
Kondo destruction \cite{Si2001,Colemanetal,senthil2004a}.
By analogy with the phase diagram of heavy fermion
metals \cite{Pas21.1,Coleman-Nature,Kirchner2020},
the existence of the phases with large and small Fermi surfaces
\cite{paschen2004,shishido2005,park-nature06} 
opens up a regime 
of amplified 
quantum fluctuations \cite{Si2001}
for 
beyond-Landau quantum criticality and 
the accompanying strange-metal behavior
\cite{Prochaska20}.
Thus, our work provides a conceptual framework to 
address the aforementioned strange-metal behavior in the flat-band-based metals.
The connection with a Kondo lattice description is also being pursued in
moir\'{e} systems \cite{Ram2021,Song2022,Kumar2022,Guerci2022.x}. The approach taken
here is expected to shed light
on the quantum phases and their transitions in those systems.

\noindent{ {\bf\large Results}} \\ 
The lattice 
contains five sublattices marked as A,\,B,\,C,\,D,\,E (Fig.~\ref{fig:disp}A).
The model is written as 
$\mathcal{H} = \mathcal{H}_0 + \mathcal{H}_1$,
which contains the on-site Hubbard interactions, 
\ba
\mathcal{H}_1 ={U}\sum_{\rr,i} n^\eta_{\rr,i,\up}n^\eta_{\rr,i,\dn} \, ,
\label{model:Hubbard}
\ea
as well as the noninteracting Hamiltonian $\mathcal{H}_0$.
Included in $\mathcal{H}_0$ are the 
nearest-neighbor tight binding hopping parameter $t$ between two sites that are connected by a solid line in Fig.~\ref{fig:disp}A, from an $\eta$ electron  {($d_{z^2}$ orbital)} located 
at site $\rr$, sublattice $i\in\{A,B,C,D,E\}$
with spin $\sigma$, to its counterpart at 
$\rr$, $j\in\{A,B,C,D,E\}$ and spin $\sigma$;
a chemical potential $\mu$;
a potential difference $m$ between the sublattices
$C,D,E$ and the sublattices $A,B$; and 
an additional potential difference $\gamma$ between
the sublattices $D,E$ 
and 
the sublattice
$C$
(see the Methods).
To illustrate our case, we consider $t=1=m$ without loss of generality.

\begin{figure}[t!]
    \centering
    \includegraphics[width=1.0\textwidth]{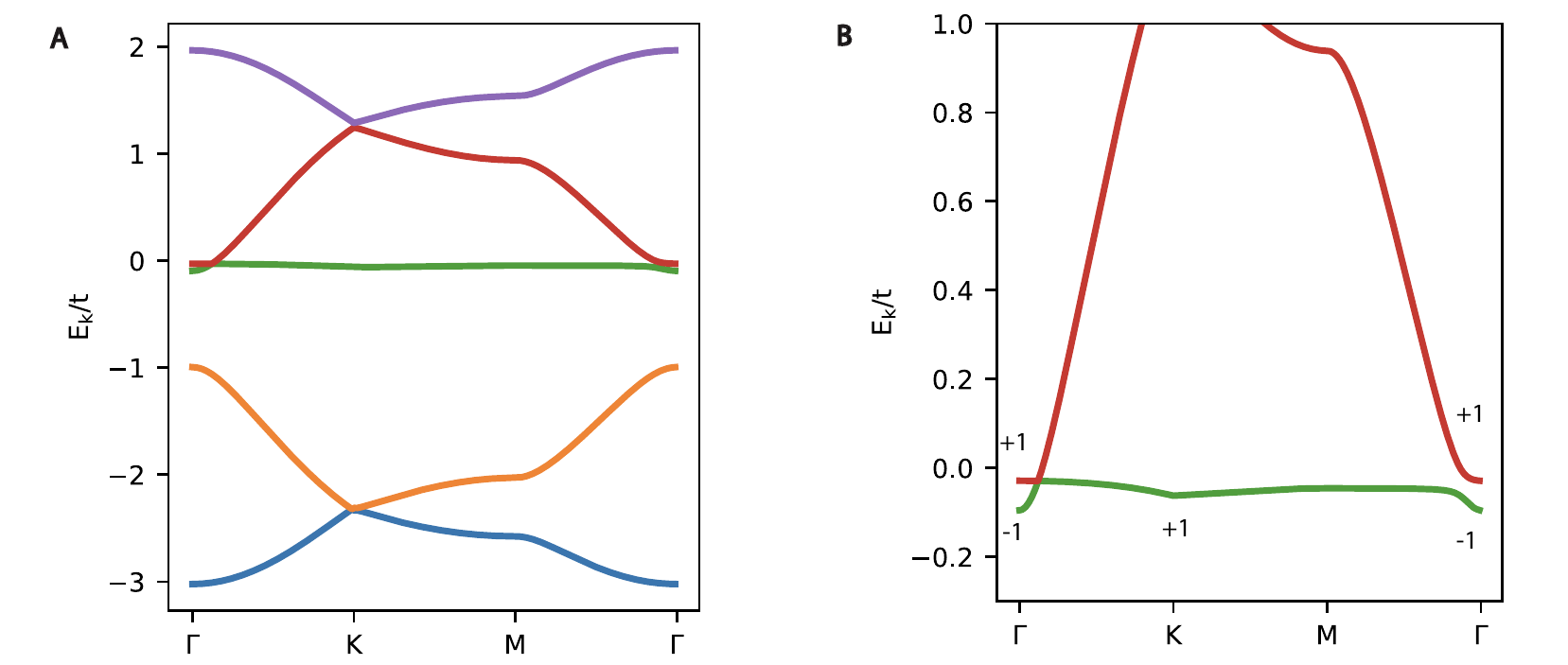}
    \caption{{\bf Noninteracting bandstructure.}
    The full bandstructure ({\bf A}) and zoomed-in bandstructure ({\bf B}) at $\gamma/t=-0.1$ 
    and $\mu/t=1.2$. In panel ({\bf B}), we label the $M_x$-symmetry eigenvalues of the bands at $M_x$-invariant momenta:  
    $\Gamma=(0,0)$ and 
    $K = (0, 4\pi/3)$.
}
\label{fig:pd}
\end{figure}

To analyze a more tractable model with a lower symmetry, we focus on the case of a non-zero $\gamma$. Here, 
a $C_{3z}$ of the $\gamma=0$ model (see the Methods)
is broken 
 and the middle band is not entirely
flat, as seen in
 figs.\,\ref{fig:pd}A, B.
 However, for small $\gamma$ (which we illustrate with the case of $\gamma=-0.1t$), 
 the middle band remains relatively flat (the bandwidth is about $0.06t$) and, thus,
 will still be referred to as a flat band. 
Near $\Gamma$, there is 
 a linear crossing between the flat and dispersive bands along 
 the $\Gamma$-$K$ line. 
 The node is located near the $\Gamma$ point and is protected by the $M_x$ symmetry. 
When a spin-orbit coupling is further included
(see the SM, Fig.\,S1E),
 such a node would be gapped out and the flat band acquires $\pm 1$ spin Chern number. 
 
 In this original lattice basis, the flat band comes from a linear superposition of the atomic states
 located in the five different sublattices. A single-atomic {-like} representation
 of the flat band is needed to make progress, which we now turn to.
 
We will consider the Hubbard interaction $U$ that is large compared to the width
of the flat band, but small compared to the width of the wide bands 
(Fig.\,\ref{fig:disp}C). Such a range is of interest to the experimentally-studied systems 
mentioned earlier.
Since there is an energy gap between the 
top three bands and the bottom two, we focus on the top three bands
that include
 the flat band.

To express a band 
in terms of exponentially localized and Kramers-doublet Wannier orbitals,
the Bloch states must have the same $M_x$-symmetry eigenvalue at all high symmetry points
\cite{cano2021band,PhysRevX.7.041069}.
In our case, the topological obstruction can be recognized 
by 
noticing
 that the different high symmetry points of the flat band have different eigenvalues of the $M_x$-symmetry. 
As shown in the SM
(band $3$ in table S1),
and illustrated in Fig.\,\ref{fig:pd}B,
 the eigenvalue changes from $+1$ at $K$ and $M'$ 
to $-1$ at $\Gamma$. Similar changes of $M_x$-symmetry eigenvalues
are 
also observed for the two wide bands (bands $1$ and $2$ in
table S1, SM).
We therefore consider a combination of Bloch states 
so that the $+1$-eigenvalue portion
of the flat band combines with the $+1$-eigenvalue portion of the wide band (band $2$, table S1, SM,
and illustrated in Fig.\,\ref{fig:pd}).

Through explicit construction, we indeed 
show the validity of the procedure. We
find
the exponentially localized Wannier functions,
which are 
 located 
on a triangular lattice (Fig.\,\ref{fig:disp}B),
with the $M_x$-symmetry eigenvalue 
being $+1$. We call such a Wannier function a $d$ orbital. 
The hopping integral of
the two neighboring $d$ orbitals is indeed weak in our construction:
It turns out to be on the order of $0.03t$.

The remaining degrees of freedom can be represented by two Wannier orbitals 
with $M_x$-symmetry eigenvalue $+1$,$-1$ respectively. 
These two orbitals mainly come from the top two dispersive bands, 
and we label them as $c$ orbitals.

We have thus succeeded in constructing a single-atomic-{like} state per unit cell to represent the flat band.
The model takes the form of an effective multi-orbital Hubbard model on a triangular lattice, which can also be viewed as an effective Anderson-lattice model.
It is expressed in terms of the $d$ and $c$ orbitals, 
containing
the kinetic term $H_0$ and the interaction term $H_I$. 

The kinetic term, $H_0$, is given in detail in the SM
It involves two types of electrons, with $d_{\RR,\sigma}^\dag$ creating a $d$-electron (electron in the orbital $d$), 
which represents the 
flat-band's degrees of freedom, at site $\RR$ with spin $\sigma$. Similarly, $c_{\RR,a,\sigma}^\dag$ 
creates a conduction $c$-electron (electron in orbital $c$), which represents the itinerant degrees of freedom, 
at site $\RR$ with spin $\sigma$ and orbital $a=1,2$.
$H_0$ contains the part $H_d$, describing the energy level ($E_d$) and their negligibly small hopping 
parameters; the part $H_c$, describing the inter-$c$-electron hopping parameters,
$ t^c_{\RR-\RR',aa'
,\sigma}$ and the energy levels $E_a$ ($a=1,2$); and the part $H_{dc}$,
the $d$-$c$ hybridization term with matrix element $V_{\RR-\RR',a,\sigma}$.
These parameters are specified in the SM.
As is seen there,
the dominant hybridization is with one of the two conduction electron bands, $a=1$.
The corresponding hybridization is off-site because $c_{\RR,1,\sigma}$ is mirror odd.

The  most important
interactions
here
include the Hubbard interactions of the $d$ electrons ($H_u$),  
and the 
Hund's coupling between the $d$ and $c$  electrons ($H_{Hund}$). 
The interactions are labeled by $u$ and
$J_{1,2}$, respectively. The specific forms 
of the interactions are given in the SM.

In the limit of $u\gg |t_{\RR-\RR'}^d|$, the Hubbard interactions suppress 
the charge fluctuations of the $d$ electrons and turn them into quantum spins. Correspondingly, the effective model acquires a representation in terms of a Kondo lattice Hamiltonian on the same triangular lattice (see the Methods). It contains two Kondo couplings $J_{K,a}$ ($a=1,2$) to two conduction $c-$electron bands and an inter-moment exchange coupling $J^H$ between the $d$-spins. Both couplings are controlled by the parameter
\ba 
\widetilde{u} = 4u\phi_0/3 \, ,
\label{eq:tilde-u}
\ea
with $\phi_0$ being 
of order unity (see the Methods and the SM, Sec.~D).

For convenience in notation, 
we now treat 
$\widetilde{u}$ of Eq.\,\ref{eq:tilde-u} as our 
tuning parameter (We note, however, that tuning $\widetilde{u}$ is equivalent to
varying the original Hubbard interaction $U$.).
At smaller $\widetilde{u}$,
the Kondo effect dominates. Increasing $\widetilde{u}$
enhances the relative effect of the antiferromagnetic exchange interactions between the local moments, which
favors the correlations between the local moments and is detrimental to the development of the Kondo effect.

We have analyzed this competition through a set of saddle-point equations that are realized in a large-$N$ limit (see the Methods).
We utilize a pseudo-fermion representation of the spin
and solve the saddle-point equations in terms of 
the field $\zeta_{\RR,\RR'a}$, which 
represents the hybridization of the Kondo-driven
composite fermions and conduction-$c$ fermions,
and $\chi_{\RR,\RR'}$, which characterizes the inter-moment spin singlets \cite{Aff1988}.

\begin{figure}[t!]
    \centering
    \includegraphics[width=0.8\textwidth]{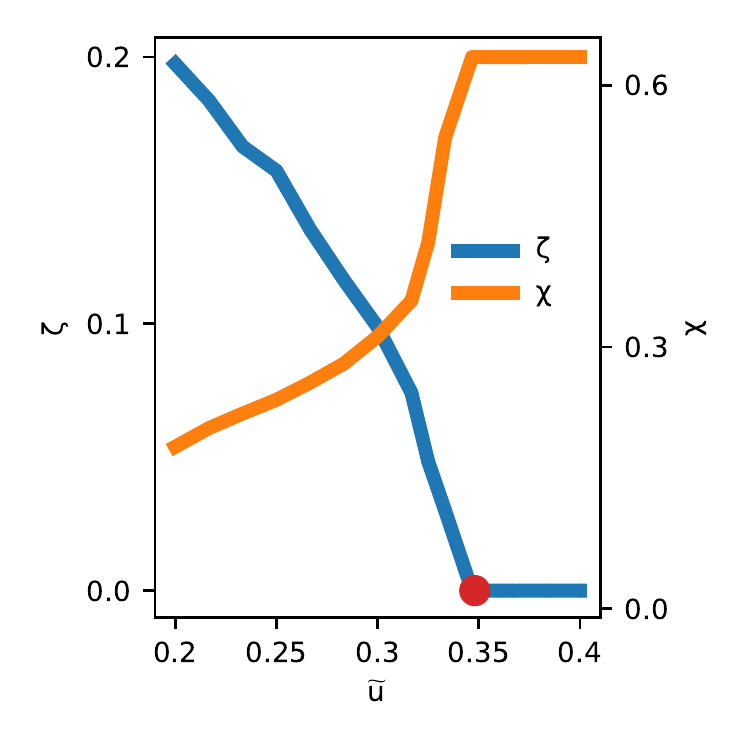}
    \caption{ {\bf The zero-temperature phase diagram calculated in the large-$N$ limit.}
    As the interaction $\widetilde{u}$ is increased, $\eta$ gradually decreases and goes to zero 
    at $\widetilde{u}=0.35$ (marked by the red dot), which signals a continuous orbital-selective Mott
    transition 
 	between a Kondo-driven 
	and Kondo-destroyed phases, with a large and a small Fermi surface respectively.
	}
    \label{fig:phase_diagram}
\end{figure}

\begin{figure}[t!]
    \centering
    \includegraphics[width=0.95\textwidth]{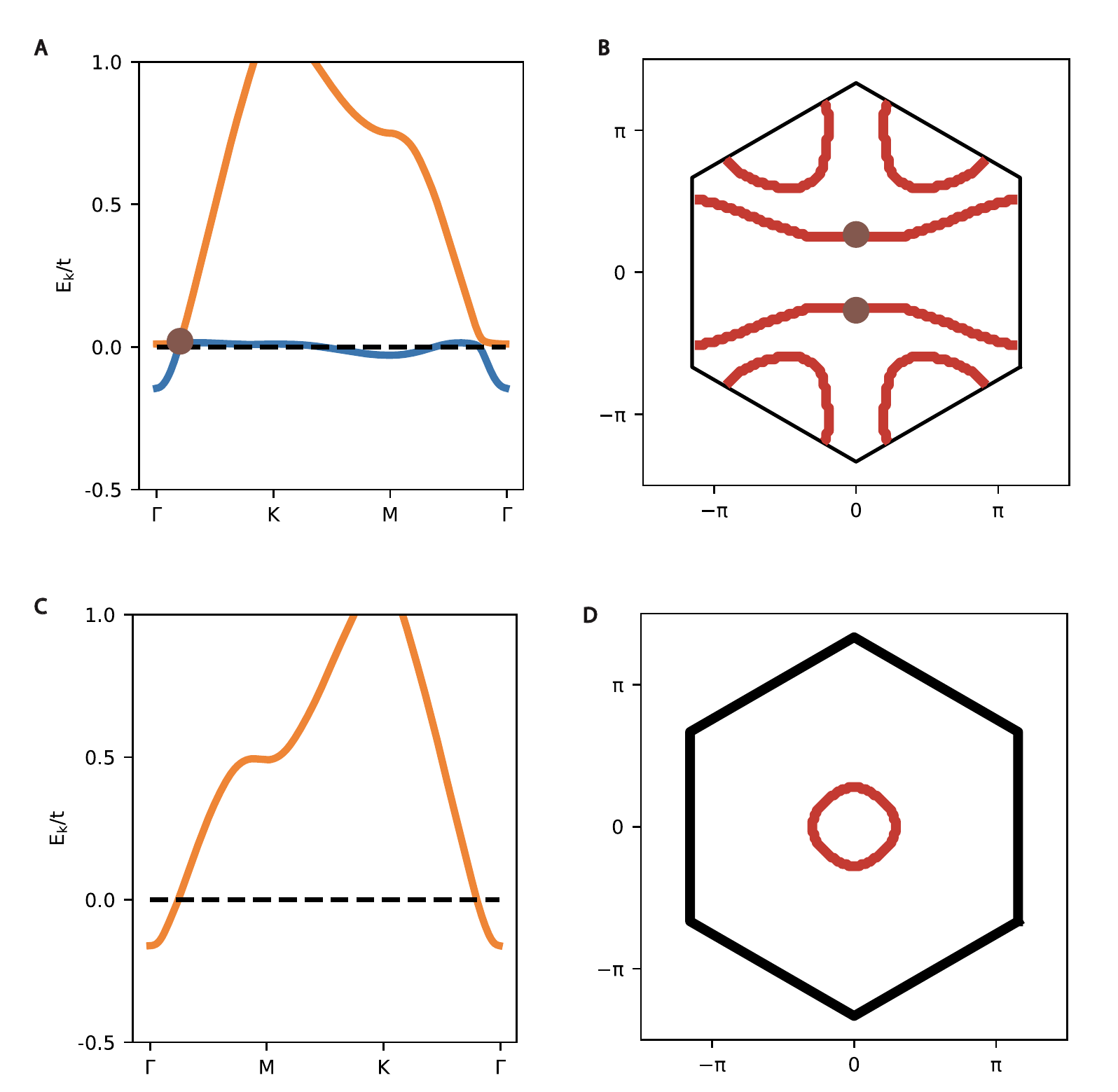}
    \caption{ {\bf Quantum phases with large and small Fermi surfaces.}
    Bandstructure (\textbf{A}) and the Fermi surface (\textbf{B}) in the phase with a large Fermi surface. Shown here 
    are the results at $\tilde{u} = 0.27$. For a larger value of $\tilde{u} = 0.35$, the corresponding
    bandstructure (\textbf{C}) and the Fermi surface (\textbf{D}) are shown for the phase with a small Fermi surface.
    In panel (\textbf{A}) and panel (\textbf{B}), the brown dot marks the Dirac nodes. 
    The dashed line in panel (\textbf{A}) and its counterpart in panel (\textbf{C}) mark the Femi energy. 
    The red lines in panel (\textbf{B}) and panel (\textbf{D}) denote the Fermi surface.   }
\label{fig:fs}
\end{figure}

Fig.~\ref{fig:phase_diagram} shows the resulting phase diagram.
At small $\tilde{u}$, we realize a  {topological} heavy Fermi liquid phase via the condensation 
of the hybridization field $\zeta$.
The latter converts the local moments into composite fermions that are represented by the 
${f}$ fields, which
 hybridize with the conduction electrons.
Furthermore, the nonzero hybridization $\zeta_{\RR_1a_2}$
 attaches a $+1$ $M_x$-symmetry eigenvalue to the
 composite fermion $f$ field,
which is the opposite to the eigenvalue of the 
dominantly
 hybridizing conducting $c_1$ band (see the SM, {Sec.~G}).
The  {correlated} heavy bands  {are topological containing} a Dirac node,
as shown in Fig.\,\ref{fig:fs}A. 
The node is protected by $M_x$ and the $SU(2)$ symmetry: 
Because of the formation of the hybridized bands, the Fermi surface encloses the electrons residing 
on both the conduction and flat bands~\cite{Piv2004,Osh2000}. This is the large Fermi surface, which is shown
in Fig.\,\ref{fig:fs}B.

Increasing $\widetilde{u}$ leads to a Kondo destruction \cite{Si2001,Colemanetal,senthil2004a}: 
The local moments couple to each other, and they no longer form a Kondo singlet with the spins 
of the conduction electrons. Consequently, the Fermi surface is derived from the dispersion 
of the conduction electrons, plotted in Fig.\,\ref{fig:fs}C.
 It 
  is shown
in Fig.\,\ref{fig:fs}D, and 
is small in the sense that the Fermi surface
encloses only the conduction electrons.
The frustrated nature of the effective lattice (Fig.\,\ref{fig:disp}B) has led to the Kondo-destroyed phase
via the bond variables $\chi_{\RR,\RR'}$.
However, the physical mechanism leading to the Kondo destruction is more 
universal, as characterized by a global phase diagram for the competition between 
the Kondo and RKKY/Heisenberg interactions \cite{Pas21.1}. 

\noindent{ {\bf\large Discussion}} \\ 
Most importantly, we have demonstrated the first realization of orbital-selective Mott transition in models of topological flat bands in the presence of coupling with wide bands. The large-Fermi surface state provides a proper description for the development of the strongly correlated $d$-electron quasiparticles. In turn, this sets the stage for the transition into the small-Fermi surface phase, which represents the destruction of the correlated quasiparticles. This orbital-selective Mott transition is in the same universality class as the 
Kondo-destruction quantum criticality of heavy fermion metals
\cite{Hu-qcm2022,Si2001,Colemanetal,senthil2004a, hu2022quantum,hu2022extended},
associated with which 
strange metallicity develops. Our theoretical description, thus, provides the conceptual framework to understand
the recently emerged experimental evidence for 
strange metal behavior in correlated
kagome metals \cite{Ye2021.x,Ekahana2021.x}.

We close with several remarks on the generality and implications of our results. Firstly, our Wannier construction procedure 
works equally well for the case with a spin-orbit coupling. 
Accordingly, our Kondo-lattice construction and
 the results for the quasiparticle formation and destruction readily apply in this case.
Secondly, by emphasizing the interplay and coupling between the flat and wide bands, 
 our work brings out a analogy between the correlated quantum materials that host 
 topological flat-bands \cite{Checkelsky2018,Yao18.1x,Comin2020-2,Dai2022,Yin2022,Setty2022}
 and moir\'{e} systems \cite{Ram2021,Song2022,Kumar2022,Guerci2022.x},
  in which the strange-metal behavior has also been indicated \cite{Jao22}. Compared to the 
  topology of the graphene-based moir\'{e} flat bands \cite{Song2022,Po19.2}, the bands in our 
  case share the traits of being topological
  but have the distinction of being much simpler.
  This simplification represents a crucial advantage in
allowing
   us to provide a proof-of-principle demonstration of both the emergence and destruction of the proper quasiparticles 
  in systems with coupled topological flat-wide bands through the Kondo route. Our explicit construction
  of the Kondo lattice also sets the stage for the study of a global quantum phase diagram \cite{Pas21.1}
   in such topological 
  flat-band-anchored systems.
   {Thirdly, the topological nature of the flat band 
  makes our Kondo lattice topological. This topological feature distinguishes our current model from the conventional Kondo lattice models where the bands are usually topologically trivial.
  This topological feature naturally makes the heavy-Fermi liquid phase 
  topologically non-trivial and featuring 
  a Dirac node.
  Beyond this, at the quantum critical point and in the non-Fermi-liquid region, the incoherent electronic excitations could also show non-trivial topology. However any such topological feature goes beyond the quasiparticle description and 
  are left for future studies.}

 {
We now discuss the stability of the quantum critical point and the corresponding non-Fermi-liquid behavior. First, at finite $N$, the Kondo destruction 
part of the phase diagram could contain a 
magnetically ordered phase. The stability of the Kondo destruction quantum critical point and the corresponding non-Fermi-liquid behavior 
remain robust, as has been 
shown in the 
Kondo lattice model~\cite{hu2022quantum,hu2022extended}. Second, with a nonzero
spin-orbit coupling, the $SU(2)$ spin symmetry will be broken. However, the existence of the Kondo destruction critical behavior relies on the competition of local momentum fluctuations and Kondo effect, instead of the specific symmetry of the local momentum~\cite{hu2022quantum,hu2022extended,PhysRevLett.91.156404}. Therefore,
the Kondo destruction quantum critical point and the non-Fermi-liquid behavior that we found here are robust.
 }  

In conclusion, we have advanced a realistic model  to study the effect of local Coulomb repulsion
for a system of coupled topological flat and wide bands. 
By constructing exponentially localized and Kramers-doublet Wannier functions 
for these bands, we are able to formulate a Kondo lattice description.
This has allowed us to provide the first demonstration of an orbital-selective Mott transition in any system of coupled 
topological flat and wide bands. 
The orbital-selective Mott transition provides 
a characterization for both the development and destruction of quasiparticles, 
leading to quantum phases with
large and small Fermi 
surfaces. Our work provides a conceptual framework to describe
the amplified quantum fluctuations of multi-band systems
with
 topological flat bands,
sets the stage to understand the 
strange-metal 
properties that are emerging
in kagome metals and other flat-band systems,
and uncovers a linkage between these systems and 
both the $f$- and $d$-electron-based correlated 
bulk materials.

\vskip 1 cm

\noindent{  {\bf\large Materials $\&$ Methods}} \\ 
The Hubbard model,
$\mathcal{H} = \mathcal{H}_0 + \mathcal{H}_1$,
is defined on a variant of the kagome lattice defined in Fig.~\ref{fig:disp}A),
with the onsite interaction term, $\mathcal{H}_1$ given in Eq.\,\ref{model:Hubbard}.
The noninteracting Hamiltonian
 is written 
 as follows,
\begin{eqnarray} 
\mathcal{H}_{0} &=& \sum_{\langle {\mathbf{r}},{\mathbf{r}}'\rangle,i,j,\sigma} t \eta_{{\mathbf{r}},i,\sigma}^\dag \eta_{{\mathbf{r}},j,\sigma}-\mu \sum_{{\mathbf{r}},i,\sigma}\eta_{{\mathbf{r}},i,\sigma}^\dag \eta_{{\mathbf{r}},i,\sigma} \nonumber \\
&&+ \sum_{{\mathbf{r}},\sigma,i 
\in \{C,D,E\} }m \eta_{{\mathbf{r}},i,\sigma}^\dag \eta_{{\mathbf{r}},i,\sigma} + \sum_{{\mathbf{r}},\sigma,i
 \in \{D,E\} }\gamma \eta_{{\mathbf{r}},i,\sigma}^\dag \eta_{{\mathbf{r}},i,\sigma}\nonumber \, .\\
\end{eqnarray}
Here, $\eta_{{\mathbf{r}},i,\sigma}^\dag$ creates an electron at site ${\mathbf{r}}$, 
sublattice $i\in\{A,B,C,D,E\}$ with spin $\sigma$. 

At $\gamma=0$, the system has a $C_{3z}$ rotational symmetry, an $M_x$ mirror symmetry 
and also an $SU(2)$ spin rotational symmetry. This leads to a purely flat band as shown in
the supplementary material (SM) (Fig.\,S1B).
There is a quadratic band touching between the 
flat and a dispersive band at 
 the center of the Brillouin zone, $\Gamma=(0,0)$. 
 The crossing is protected by both the $M_x$ and $C_{3z}$ symmetry.

To analyze a more tractable model with a lower symmetry, we focus on the case of a non-zero $\gamma$,
as described in the main text.

In the limit of $u\gg |t_{\mathbf{R}-\mathbf{R}'}^d|$, 
the charge fluctuations of the $d$ electrons
are suppressed and these electrons are 
turned into quantum spins. 
By integrating out the high energy degrees of freedom [in the presence of the Hund's coupling \cite{Ding19.2,Meetei13}],
we reach a Kondo-Heisenberg model with the Hund's coupling. 
The effective Kondo-lattice Hamiltonian on the triangular lattice is
\begin{eqnarray} 
H_{KH} &=& H_c +H_{H} +H_{K}+ H_{Hund} \nonumber \\  
H_H&=&\sum_{\mathbf{R},\mathbf{R}'}J^H_{\mathbf{R},\mathbf{R}'}{{\vec{S}}}_{\mathbf{R}} \cdot 
{{\vec{S}}}_{\mathbf{R}'} \nonumber \\
H_K &=& \sum_{\mathbf{R},\mathbf{R}_1,a_1,\mathbf{R}_2,a_2}J^K_{\mathbf{R},\mathbf{R}_1 a_1,\mathbf{R}_2 a_2}{{\vec{S}}}_{\mathbf{R}} \cdot c_{\mathbf{R}_1,a_1}^\dag \vec{\sigma} c_{\mathbf{R}_2,a_2}
\, .
\nonumber \\
\end{eqnarray} 
Here, 
 $S_\mathbf{R}$ is the emergent spin-$1/2$ local moment formed by the localized $d$ electrons, 
and $ {{\vec{S}}}^c$ is the corresponding spin of the conduction electrons.
In the large $u$ limit, the hoppings of $d$ electrons induce a Heisenberg interaction 
of strength $J^H_{\mathbf{R},\mathbf{R}'} = 2|t_{{\mathbf{R}},{\mathbf{R}}'}^d|^2/ \widetilde{u}$ where, as
$
\widetilde{u}$ is specified by Eq.\,\ref{eq:tilde-u}.
The non-local hybridization terms between the $d$ and $c$ electrons 
lead to non-local Kondo couplings of strength 
$J^K_{{\mathbf{R}},{\mathbf{R}}_1,{\mathbf{R}}_2,a_1,a_2} =4V_{{\mathbf{R}}-{\mathbf{R}}_2,a_2}^* V_{{\mathbf{R}}-{\mathbf{R}}_1,a_1} /\widetilde{u}$. 
Here, the local moment of the $d$ electron is Kondo coupled to a spin operator $ c_{{\mathbf{R}}_1 a_1}^\dag \mathbf{\sigma} c_{{\mathbf{R}}_2a_2}$ 
that are formed by two electron operators from the sites $R_1,R_2$ and the 
orbitals $a_1,a_2$ respectively. 
Finally, the Hamiltonians for the $c$ electrons and for the Hund's coupling remain unchanged from those given
in the effective multi-orbital Hubbard (i.e., the effective Anderson-lattice) model. In particular, the Hund's coupling acts
between the $d$ and $c$ electrons of the same site
(see the SM Sec.~C).

To analyze the competition between the inter-moment and Kondo exchange couplings,
 we introduce the pseudo-fermion representation of the spin 
operators 
${{\vec{S}}}_{{\mathbf{R}}} = \frac{1}{2}\sum_{\sigma,\sigma'}f_{{\mathbf{R}},\sigma}^\dag \vec{\sigma}_{\sigma,\sigma'} f_{{\mathbf{R}},\sigma'}$ and 
solve the model in the large-$N$ limit [with a generalization of $SU(2)$ to $SU(N)$ and a suitable rescaling of the coupling
constants in terms of $1/N$; see the SM Sec.~F].
The ground state is characterized by the bond fields $\chi_{{\mathbf{R}},{\mathbf{R}}'}$ 
 \cite{Aff1988}
 and the hybridization fields $\zeta_{{\mathbf{R}},{\mathbf{R}}'a}$:
\begin{eqnarray}
&&\chi_{{\mathbf{R}},{\mathbf{R}}'}=\frac{1}{N}\sum_{\sigma}\langle f_{{\mathbf{R}},\sigma}^\dag f_{{\mathbf{R}}',\sigma} \rangle \nonumber \\
&&\zeta_{{\mathbf{R}},{\mathbf{R}}'a}=\frac{1}{N}\sum_{\sigma}\langle f_{{\mathbf{R}},\sigma}^\dag c_{{\mathbf{R}}',a,\sigma} \rangle\, ,
\end{eqnarray} 
and their maximum amplitudes $\zeta =$max$ {}_{\mathbf{R},\mathbf{R}',a} \{ |\zeta_{\mathbf{R},\mathbf{R}'a}|\}$ 
and $\chi =$max${}_{{\mathbf{R}},\mathbf{R}'} \{ |\chi_{\mathbf{R},\mathbf{R}'}| \}$. 
The results from solving the saddle-point equations have been given in the main text. 
 {The Kondo destruction is captured by the suppression of the hybridization fields $\zeta$, which appears 
not only in
the $SU(2)$-symmetric setting but also 
in the cases with
spin anisotropy ~\cite{PhysRevLett.91.156404, hu2022extended,hu2022quantum}.
}

\bibliographystyle{Science}
\bibliography{hfflat}

\noindent{\bf Acknowledgments:}
We thank 
Lei Chen, Silke Paschen, Chandan Setty,  Ming Yi and, especially,
Gabriel Aeppli for useful discussions. 

\noindent {{\bf Funding:}}
The work was primarily supported by the 
U.S. DOE, BES, under Award No.\ DE-SC0018197
and additionally supported by
the Robert A.\ Welch Foundation Grant No.\ C-1411.
The majority of the computational calculations have been performed on the
Shared University Grid at Rice funded by NSF under Grant EIA-0216467, a partnership between
Rice University, Sun Microsystems, and Sigma Solutions, Inc., the Big-Data Private-Cloud 
Research Cyberinfrastructure MRI-award funded by NSF under Grant No. CNS-1338099 and by
Rice University, and the Extreme Science and Engineering Discovery Environment (XSEDE) by NSF
under Grant No. DMR160057. 
The work of Q.S.\ was performed in part at the Aspen Center for Physics, 
which is supported by the NSF grant No.\ PHY-1607611.

\noindent$^{\ast}$To whom correspondence should be addressed; E-mail:  qmsi@rice.edu.

\newpage

\clearpage

\newpage

\clearpage

\newpage

\clearpage

\newpage

\setcounter{figure}{0}
\setcounter{equation}{0}
\makeatletter
\renewcommand{\thefigure}{S\@arabic\c@figure}
\renewcommand{\theequation}{S\arabic{equation}}
\renewcommand\thetable{S\@arabic\c@table}

\section*{Supplementary Material}

\subsection*{A. Band topology and Wannier construction}
\label{sec:band_top}
\begin{figure}
    \centering
    \includegraphics[width=1\textwidth]{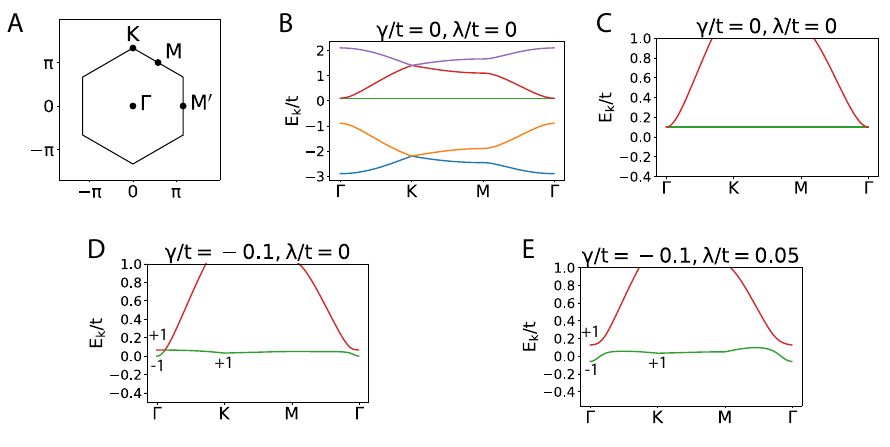}
    \caption{{\bf Brillouin zone and the noninteracting bandstructure.}
   The first Brillouin zone ({\bf A}). Full band structure ({\bf B}) and zoomed-in band structure ({\bf C}) 
   at $\gamma=\lambda=0$ with a completely flat band near the Fermi energy. 
   ({\bf D}) Band structure at non-zero $\gamma$ and $\lambda=0$. 
   We observe a relatively flat band and a Dirac node between the flat and wide bands. ({\bf E})
   Band structure at a non-zero $\gamma$ and a non-zero $\lambda$, the Dirac node is gapped by 
   the spin-orbit 
   coupling and the flat band acquires $\pm 1$ Chern number. In ({\bf D}) and ({\bf E}), 
   we label the $M_x$ eigenvalues of the relevant bands at the high symmetry points.}
    \label{fig:band}
\end{figure}

The space group is generated by the translational symmetry and the mirror $M_x$ symmetry. 
At $\gamma=0$, the model has an additional $C_{3z}$ symmetry. 
We observe a completely flat band and a quadratic band touching between
the flat and wide bands, as shown in 
Figs.~\ref{fig:band}B,C. By introducing a non-zero $\gamma$, 
the flat band is no longer completely flat, but its bandwidth remains relatively narrow at small $\gamma$. 
Near $\Gamma$ there is a linear crossing (Dirac node) between the flat band and wide band as shown in 
fig.~\ref{fig:band}D. Such Dirac node is protected by the $M_x$ and $SU(2)$ spin symmetry. 
If we introduce spin-orbit coupling that breaks the $SU(2)$ spin symmetry, the Dirac node 
will be gapped out and the flat band acquires $\pm 1$ Chern number. In fig.~\ref{fig:band}E, 
we show the band structure of the following spin-orbit coupling 
\ba 
H_{soc} =  \sum_{\kk,\sigma} i\lambda \sigma \bigg[  \eta_{\kk,C,\sigma}^\dag \eta_{\kk,D, \sigma} + \eta_{\kk,D,\sigma}^\dag \eta_{\kk,E,\sigma} +\eta_{\kk,E,\sigma}^\dag \eta_{\kk,C,\sigma} \bigg] +\text{h.c.}
\ea 

We can also observe the non-trivial topology by calculating the $M_x$ eigenvalue of each band at high symmetry points,  {where $\eta^\dag$ creates an electron  {with the $d_{z^2}$ orbital} that is even under $M_x$ transformation}.
We focus on the case of $\gamma \ne 0 $ and show the result in table~\ref{tab:sym} and also in fig.~\ref{fig:band}D,E. We note that, at $\gamma\ne 0$, each band forms a one-dimensional (1d) irreducible representation of the little group at each high-symmetry point, which is characterized by the $M_x$ eigenvalue. However, in the case of 
$\gamma=0$, we have both $C_{3z}$ and $M_x$ symmetries, and therefore, a two-dimensional (2d) irreducible representation could be formed by the bands at high symmetry points. The case of $\gamma\ne 0$ simplifies the construction of the Wannier orbitals but still keeps the non-trivial topology of the band structure. 
 {At $\gamma \ne 0$, the flat band is non-longer perfectly flat. However, as long as 
$|\gamma| \lesssim 0.2t$, 
the middle bands still remain relatively flat and the consideration of Wannier construction in this section still holds.}

The high symmetry points are $\Gamma=(0,0),M'=(2\pi/\sqrt{3},0), K = (0,4\pi/3)$ and we label the bands from the top to 
the 
bottom as band $1,2,3,4,5$, with band $3$ being the flat band. We note that a non-zero but small $\lambda$ will not change the $M_x$ eigenvalue of each band. If we only consider the flat band (band 3), its $M_x$ eigenvalues at $\Gamma,M,K$ are $-1,+1,+1$ respectively, which do not admit a Wannier-orbital representation 
that preserves all the symmetry. 
However, by taking band 2 at the $\Gamma$ point and band $3$ at the $M'$ and $K$ points, we have the 
$M_x$ eigenvalue $+1,+1,+1$ which can be represented by an $M_x$-even Wannier orbital. 
This indicates that we can construct the Wannier orbitals by combining bands 2 and 3. A related procedure has been taken for moir\'{e} systems \cite{Song2022}. As we show later, the constructed Wannier orbital would have a large overlap with the flat band and can be used to represent the flat-band degrees of freedom. After removing the $+1,+1,+1$ eigenvalues at $\Gamma,M',K$, the remaining eigenvalues are $+1,-1$ for $\Gamma$ point, $+1,-1$ for $M'$ point and  $+1,-1$ for $K$ point. These combinations admit two Wannier orbitals that are mirror even and mirror odd respectively. These two Wannier orbitals mainly describe wide-band degrees of freedom.

We now
construct the Wannier orbitals for the case of $\gamma=-0.1,\lambda=0$. However, 
the same procedure applies to the case with a small but non-zero $\lambda$. 
We introduce the following three trial wavefunctions (six trial wavefunctions if we account for
the spin degeneracy) and construct the Wannier orbitals of the top three bands via the projection method:
\begin{eqnarray*}
&&|\text{Trial};d,\RR,\sigma\rangle = \sum_{\rr,i} e^{- \frac{(\rr-\RR)^2}{2r_0^2}}\alpha^f_i \eta_{\rr,i,\sigma}|0\rangle\\
&&|\text{Trial};c,1,\RR,\sigma\rangle = \sum_{\rr,i} e^{- \frac{(\rr-\RR)^2}{2r_0^2}}r_x \eta_{\rr,i,\sigma}|0\rangle 
\\
&&|\text{Trial};c,2,\RR,\sigma\rangle = \sum_{\rr,i} e^{- \frac{(\rr-\RR)^2}{2r_0^2}}r_y \eta_{\rr,i,\sigma}|0\rangle 
\label{eq:trial_state}
\end{eqnarray*}
where $r_0 =0.25a_0$, with $a_0$ being the distance between the two nearest atoms,
 describes the decaying rate. $\alpha^d_{A,B,C,D,E} ( = 0.5,0.5,1.4,1,1)$ captures the imbalance of weight distributions between different sublattices. After the projection procedure, we find three exponentially-localized Wannier orbitals $|d,\RR,\sigma\rangle, |c,\RR,1,\sigma\rangle, |c,\RR,2,\sigma\rangle$, which come
  from the three trial wavefunctions respectively. 
  We then define the electron operators in the Wannier basis via
\begin{eqnarray*} 
d_{\RR,\sigma}^\dag |0\rangle = |d,\RR,\sigma\rangle \quad,\quad 
c_{\RR,1,\sigma}^\dag |0\rangle = |c,\RR,1,\sigma\rangle \quad,\quad 
c_{\RR,2,\sigma}^\dag |0\rangle = |c,\RR,2,\sigma\rangle \,. 
\end{eqnarray*}
By construction, both $d_{\RR,\sigma}$ and $c_{\RR,2,\sigma}$ are even under $M_x$ and $c_{\RR,1,\sigma}$ is 
odd under $M_x$. 

In addition, we also calculate the orbital weight of each band, which characterizes the overlapping between each band 
and each Wannier orbital. The overlappings between the Wannier orbitals $d_{\kk,\sigma},c_{\kk,1,\sigma}$ 
and $c_{\kk,2,\sigma}$ and the $i$-th band are defined as 
\begin{eqnarray*}
&&W^{d}_i = \frac{1}{2N}\sum_{\kk} \sum_\sigma|\langle u_{\kk,i,\sigma}| d,\kk,\sigma \rangle |^2 \\
&&W^{c,1}_i = \frac{1}{2N}\sum_{\kk} \sum_\sigma|\langle u_{\kk,i,\sigma}| c,\kk,1,\sigma \rangle |^2 \\
&&W^{c,2}_i = \frac{1}{2N}\sum_{\kk} \sum_\sigma|\langle u_{\kk,i,\sigma}| c,\kk,2,\sigma \rangle |^2 
\end{eqnarray*} 
where $|u_{\kk,i,\sigma}\rangle$ is the Bloch function of the $i$-th band with spin index $\sigma$, $|d,\kk,\sigma,\rangle ,|c,\kk,1,\sigma \rangle , |c,\kk,2,\sigma \rangle $ are the Fourier transformation of $|d,\RR,\sigma \rangle ,|c,\RR,1,\sigma \rangle ,|c,\RR,2,\sigma \rangle$ respectively. Because of the $SU(2)$ spin symmetry, we average over 
the contributions between the spin up and spin down. From numerical calculations, we have 
\begin{eqnarray}
&&W^d_1 = 0\% \quad,\quad W^d_2 =12\%  \quad,\quad W^d_3 = 88\% \quad,\quad W^d_4 = 0\%\quad,\quad W^d_5 = 0\%  \nonumber \\
&&W^{c,1}_1 =34\% \quad,\quad W^{c,1}_2 =57\% \quad,\quad W^{c,1}_3 = 9\%
\quad,\quad W^{c,1}_4 = 0\%\quad,\quad W^{c,1}_5 = 0\%\nonumber  
\\
&&W^{c,2}_1 =66\% \quad,\quad W^{c,2}_2 =31\% \quad,\quad W^{c,2}_3 =3\% \quad,\quad 
 W^{c,2}_4 = 0\%\quad,\quad W^{c,2}_5 = 0 \nonumber \%
 \label{eq:weight_non_soc}
\end{eqnarray}
where we label the bands from the top to the bottom as $1,2,3,4,5$; band $3$ is the flat band. 
We observe that the flat bands are mainly represented by the $d$ orbitals, and the two dispersive bands 
($1$ and $2$) are mainly captured by the two $c$ orbitals.

\begin{figure}
    \centering
    \includegraphics[width=0.5\textheight]{ 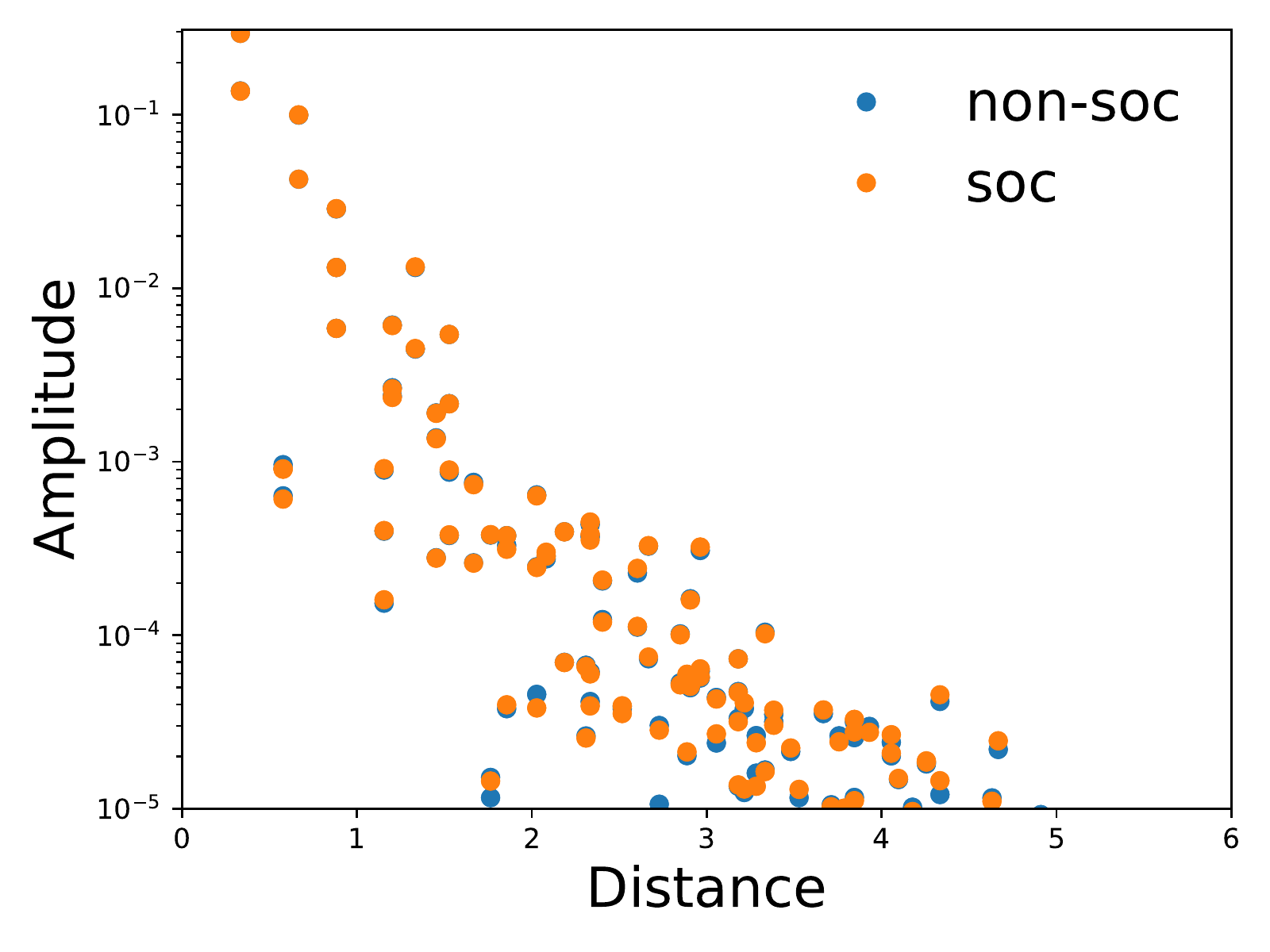 }
    \caption{   { {\bf Decaying of the wavefunction amplitudes of the Wannier function.}}  {The Wannier functions of the $d$ electrons (with non-zero SOC and zero SOC) decay exponentially. The decaying of the two cases is similar with only a small deviation at long distances, since the two Wannier orbitals are adiabatically connected with each other.} }
    \label{fig:weight}
\end{figure}

 {Our Wannier construction equally applies for the case with spin-orbit coupling. By implementing the same trial state as described in Eq.~\ref{eq:trial_state}, we are able to construct exponentially-localized Wannier orbitals without breaking any symmetry of the Hamiltonian. The resulting Wannier function at relatively small but still sizeable spin-orbit coupling $(\lambda/t =0.05)$ is adiabatically connected to the Wannier functions at zero spin-orbit coupling. Similarly as for the case  without spin-orbit coupling, we observe that the flat bands are mainly represented by the $d$ orbitals, and the two dispersive bands (1 and 2) are mainly captured by the two $c$ orbitals. The corresponding orbital weights are listed below. 
}
 { 
\begin{eqnarray}
&&W^d_1 = 0\% \quad,\quad W^d_2 =12\% \quad,\quad W^d_3 = 88\% \quad,\quad W^d_4 = 0\%\quad,\quad W^d_5 = 0\% \nonumber \\
&&W^{c,1}_1 =35\% \quad,\quad W^{c,1}_2 =57\% \quad,\quad W^{c,1}_3 = 8\%
\quad,\quad W^{c,1}_4 = 0\%\quad,\quad W^{c,1}_5 = 0 \nonumber \%
\\
&&W^{c,2}_1 =65\% \quad,\quad W^{c,2}_2 =31\% \quad,\quad W^{c,2}_3 =4\% \quad,\quad 
 W^{c,2}_4 = 0\%\quad,\quad W^{c,2}_5 = 0 \% \nonumber \\ 
\end{eqnarray}
where we observe, even with the nonzero spin-orbit coupling, the flat band can still be faithfully represented by a localized $d$ electron operators (where the overlapping between flat band and $d$ electron operator is $88\%$. In fig.~\ref{fig:weight}, we plot the decaying of the Wannier functions of $d$ electrons for both cases of zero and nonzero spin-orbit coupling (SOC). We observe that, in both cases, the Wannier functions are well localized and decay exponentially. Moreover, the Wannier functions in two cases share the similar decaying patterns since a nonzero but small SOC only leads to a small change to the Wannier function.
}

 {For this work, we will focus on the model without spin-orbit coupling to simplify the calculations. However, we mention that, our results apply to the case with nonzero SOC. Even though the nonzero SOC breaks the SU(2) symmetry, phases and critical points remain robust against a nonzero but small SOC. The stability of the different phases is seen as follows. For the Kondo or heavy-Fermi-liquid phase, the small SU(2) breaking is irrelevant in the renormalization group sense and the Kondo physics has been observed experimentally and theoretically in various different systems without SU(2) symmetry~\cite{PhysRevLett.91.156404}. For the Kondo-destroyed phases and Kondo destruction quantum critical point, it has been demonstrated via both the renormalization group and quantum Monte Carlo methods that both could be stabilized in the cases without an SU(2) symmetry~\cite{PhysRevLett.91.156404, hu2022extended,hu2022quantum}. 
}

\subsection*{B. Parameters of the model}
\label{sec:para}
After constructing the Wannier orbitals, we project the hopping and interaction terms to the Wannier basis, 
which leads to an interacting model.
In this section, we provide the parameters of the interacting model. 
The parameters of the hopping and hybridization are shown in table.~\ref{tab:par1}.  
We also
determine the dominant local interactions as follows,
\ba 
&&u/U = 0.149\quad,\quad 
F_1 /U= 0.044  \quad,\quad F_2/U = 0.036 \quad,\quad J_1/U = 0.088 \quad,\quad J_2/U =0.072 \nonumber \\
\ea 

Here, the pairing hopping terms ({\it i.e.} $d^\dag d^\dag cc, c^\dag c^\dag d d$) have negligible effect,
because we focus on the interaction range where the Hubbard interaction $u$ turns the $d$ electron into a quantum spin, and are hence dropped. 
Moreover, the interactions between the $c$ electrons only renormalize the bandwidth of the $c$ electrons,
given that the bandwidth of the $c$ electrons is much larger than the interactions in the parameter region we consider; thus, the interactions between the $c$ electrons are also dropped.

\subsection*{C. Effective multi-orbital Hubbard (Anderson-lattice) model}
\label{sec:model}
The model, expressed in terms of the $d$ and $c$ orbitals, 
contains
the kinetic term $H_0$ and the interaction term $H_I$. 
The kinetic term takes the following form:
\ba 
H_0 &=&H_d +H_c +H_{dc}\nonumber \\
H_d&=& \sum_{\RR,\RR',\sigma}\bigg[t^d_{\RR-\RR',\sigma} +(E_d-\mu)\delta_{\RR,\RR'}\bigg]d_{\RR,\sigma}^\dag d_{\RR',\sigma} \nonumber \\
H_c&=&
\sum_{\RR,\RR',a,a',\sigma}\bigg[ t^c_{\RR-\RR',aa'
,\sigma}-\mu\delta_{\RR,\RR'}\delta_{a,a'}\bigg] 
c_{\RR,a,\sigma}^\dag c_{\RR',a',\sigma}
+\sum_{\RR,a,\sigma}E_a c_{\RR,a,\sigma}^\dag c_{\RR,a,\sigma} \nonumber 
\\
H_{dc}&=& 
\sum_{\RR,\RR',a,\sigma}\bigg[ V_{\RR-\RR',a,\sigma}d_{\RR,\sigma}^\dag c_{\RR',a',\sigma} +\text{h.c.} \bigg] 
\, .
\label{Ham:H_0-d-c-dc}
\ea 

The dominant hybridization is between $d_{\RR,a,\sigma}$ and the new conduction-electron
band $1$, $c_{\RR,1,\sigma}$, which is mirror odd; correspondingly, the hybridization is off-site.
For interactions, it suffices to keep the ones between two $d$-electrons and those between the 
$d$- and $c$-electrons. As mentioned earlier,
the interactions between the $c$-electrons are omitted; they are unimportant, being small 
 compared 
to the corresponding bandwidth.

The interactions include the Hubbard interaction 
of the $d$ electrons ($H_u$), 
the density-density interactions between the 
$d$ and $c$ electrons $(H_F)$ (which is unimportant in the local-moment regime that we focus on),
and the 
Hund's coupling between the $d$ and $c$  electrons ($H_{Hund}$). 
These interactions are labeled by $u,F_{1,2},J_{1,2}$, respectively. The full interacting part of the
Hamiltonian takes the form of
\ba 
&&H_I=H_u +H_F +H_{Hund} \nonumber \\
&&H_u=\sum_{\RR}\frac{u}{2}(n^d_{\RR}-1)^2 \quad,\quad 
H_F =\sum_{\RR,a }F_a n^f_{\RR}n^c_{\RR,a}, \nonumber\\
&&H_{Hund} =-\sum_{\RR,a} J_a \bm{S}_{\RR} \cdot \bm{S}^{c}_{\RR,a}  \, .
\label{Ham:H_I-u-F-Hund}
\ea 
Here, $\bm{S}_{\RR} = \frac{1}{2}d^\dag_{\RR} \bm{\sigma} d_{\RR}$ and $ \bm{S}^c_{\RR,a} = \frac{1}{2}c_{\RR,a}^\dag \bm{\sigma} c_{\RR,a} $ 
are the 
spin operators of the $d$- and $c$-electrons, respectively; $n^{d}_{\RR,\sigma}=d_{\RR,\sigma}^\dag d_{\RR,\sigma}
$ and $n^c_{\RR,a,\sigma} = c_{\RR,a,\sigma}^\dag c_{\RR,a,\sigma}$ are the density operators of 
the $d$ and $c$ electrons, respectively;
$n^d_{\RR} = \sum_{\sigma} n^d_{\RR,\sigma}, n^c_{\RR,a} = \sum_{\sigma} n^c_{\RR,a,\sigma}$.

\subsection*{D. Effective Kondo-lattice model} 
\label{sec:kondo}
In this section, we derive the low-energy effective model for sufficiently large values of $u$.
 The action of the original Hamiltonian reads
\begin{eqnarray*}
S =\int_\tau \sum_{\RR,\sigma}d_{\RR,\sigma}^\dag \partial_\tau d_{\RR,\sigma} + \int_\tau \sum_{\RR,a,\sigma}c_{\RR,a,\sigma}^\dag \partial_\tau c_{\RR,a,\sigma} + \int_\tau (H_d+H_c+H_{dc} +H_U+H_F +H_{Hund}) d\tau \,. 
\end{eqnarray*}
We perform a Hubbard-Stratonovich transformation to decouple the Hubbard interaction and Hund's coupling term
\begin{eqnarray*} 
Z&=& \int_{d,d^\dag, c,c^\dag} e^{-S} \\
&=&\int_{d,d^\dag,c,c^\dag, \phi}\exp\bigg\{ 
-\int_\tau \sum_{\RR,\sigma}d_{\RR,\sigma}^\dag \partial_\tau d_{\RR,\sigma} - \int_\tau \sum_{\RR,a,\sigma}c_{\RR,a,\sigma}^\dag \partial_\tau c_{\RR,s,\sigma} - \int_\tau (H_f+H_c+H_{dc} +H_F) \\
&& -\int_\tau \sum_{\RR}\frac{2u}{3}\bm{\phi}_{\RR}\cdot \bm{\phi}_{\RR}  + \int_\tau\frac{4u}{3}\sum_{\RR} \bm{\phi}_\RR \cdot (\bm{S}^d_\RR +\sum_a \frac{3J_a}{4u} \bm{S}_{\RR a}^c ) 
\bigg\} 
\end{eqnarray*}
where we have dropped the interactions between the conduction electrons. We then introduce the special unitary matrix $U_\RR$ that satisfies 
\ba 
\phi_0 U_\RR^\dag \sigma^z U_\RR = \sum_\mu \phi_\RR^\mu \sigma^\mu 
\ea 
and the new fermions operators
\ba 
\psi_{\RR,\sigma} = \sum_{\sigma'}[U_\RR^\dag]_{\sigma\sigma'}d_{\RR,\sigma'}\,.
\ea  
We parametrize $U_\RR$ with 
\ba 
U_\RR = \begin{bmatrix}
z_{\RR,1} & -z_{\RR,2}^\dag \\
z_{\RR,2} & z_{\RR,1}^\dag 
\end{bmatrix} 
\ea 
with $\sum_a z_{\RR,a}^\dag z_{\RR,a}=1$.
We then have the following action 
\begin{eqnarray*} 
S& = &S_\psi + S_B + S_t + S_V+S_F +S_c +S_\phi  \\
S_\psi& = &\int_\tau \sum_{\RR,\sigma} \psi_{\RR,\sigma}^\dag \partial_\tau \psi_{\RR\sigma} - \int_\tau\frac{4u}{3}\sum_{\RR} \phi_0 \psi_{\RR}^\dag \sigma^z \psi_{\RR} \\
S_B &=&\int_\tau \sum_{\RR} \psi_\RR^\dag(U_\RR^\dag \partial_\tau U_\RR)\psi_\RR \\
S_t &=&\int_\tau \sum_{\RR,\RR'}t_{\RR,\RR'}\psi_\RR^\dag U_\RR^\dag U_{\RR'} \psi_{\RR'} \\
S_V&=&\int_\tau \sum_{\RR,\RR',a} V_{\RR,\RR'a}\psi_{\RR}^\dag U_\RR^\dag c_{\RR',a,\sigma} \\
S_c &=& \int_\tau \sum_{\RR,a,\sigma}c_{\RR,a,\sigma}^\dag \partial_\tau c_{\RR,s,\sigma} + \int_\tau\frac{4u}{3}\sum_{\RR}\phi_0 U_\RR^\dag \sigma^z U_\RR \cdot (\sum_a \frac{3J_a}{4u} \bm{S}_{\RR a}^c ) +\int_\tau \sum_{\RR,a} F_a \sum_\sigma \psi_{\RR,\sigma}^\dag \psi_{\RR,\sigma} n_{\RR,c}^c +H_c \nonumber \\
S_\phi& =&\int_\tau \sum_{\RR} \frac{2u}{3}\bm{\phi}_\RR \cdot \bm{\phi}_{\RR}
\end{eqnarray*}
The large $U$ leads to a local moment formation, which gives $\phi_0 \ne 0 $ at the saddle point level. A non-zero $\phi_0$ will then gap out the $\psi$ fermions. We can safely integrate out the gapped $\psi$ fermion, which gives the following new action at the leading order
\begin{eqnarray*} 
&&S' = S_{Berry} +S_{Heisenberg} +S_{Kondo} + S_c\\
&&S_{Berry} =\langle S_B\rangle_0 =-\int_\tau \sum_{\RR,a} z_{\RR,a}\partial_\tau z_{\RR,a}^\dag \\
&&S_{Heisenberg} =-\frac{1}{2} \langle S_t^2 \rangle_0  = \int_\tau \sum_{\RR,\RR'}\frac{2|t_{\RR,\RR'}|^2}{4u\phi_0/3}
\bm{S}_\RR \cdot \bm{S}_{\RR'} \\
&&S_{Kondo} =-\frac{1}{2} \langle S_V^2 \rangle_0  = \int_\tau \sum_{\RR,\RR_1a_1,\RR_2a_2}\frac{4V_{\RR_,\RR_2a_2}^* V_{\RR,\RR_1a_1}}{4u\phi_0/3}
\bm{S}_\RR \cdot c_{\RR,a_1}^\dag \bm{\sigma} c_{\RR,a_2}
\end{eqnarray*} 
where for a given operator $O$, $\langle O\rangle_0 =\bigg[\int_{\psi,\psi^\dag}O 
e^{-S_\psi}\bigg] /\bigg[\int_{\psi,\psi^\dag} e^{-S_\psi}\bigg]$ 
and the spin operator $\bm{S}_\RR$ is defined as $\bm{S}_{\RR} = \frac{1}{2}z^\dag_\RR \bm{\sigma}z_\RR$, 
which is the spin moment of a $d$ electron. $S_{Berry}$ is the Berry phase term of the spin operator $\bm{S}_\RR$, $S_{Heisenberg}$ and $S_{Kondo}$ are the Heisenberg interaction and Kondo interaction term respectively.
Finally, we transform the action to the Hamiltonian and reach the Kondo-Heisenberg Hamiltonian with 
the Hund's coupling as shown in the main text.

\subsection*{E. Competition between the Heisenberg/RKKY interactions and Kondo coupling}
\label{sec:comp}
There are two emergent energy scales in the Kondo lattice model: 
the Kondo energy scale $E_K$ and the Heisenberg/RKKY scale $E_H$. 
The Kondo scale $E_K \sim D e^{-1/(\rho J^K)}$, where $J^K = \text{max}\{J^K_{\RR,\RR_1a_1,\RR_2a_2}\}$ 
is the maximum amplitude of the Kondo coupling, 
$\rho$ is the $c$-electron density of states at the Fermi energy, and $D$ is the bandwidth of the $c$ bands. 
The Heisenberg/RKKY scale $E_H$ describes the inter-moment interactions, and is 
 proportional to the $J^H =\text{max}\{J_{\RR,\RR'}^H\}$. When $E_K \gg E_{H}$, the Kondo physics dominates; the $d$ electrons are Kondo-screened by the 
$c$ electrons and a heavy Fermi liquid ensues. 
When $E_H \gg E_K$, 
the inter-moment exchange interactions 
prevail. They favor the formation of spin singlets between the $d$-spins and, thus, are detrimental to the Kondo effect. The competition between 
the two 
effects
is conveniently captured by tuning
$\widetilde{u}$. 

We remark that the effective Hund's coupling here is different from the usual atomic cases in an important way. Here, it operates between the localized $d$-electrons and the wide-band $c$ electrons. 
Thus, unlike the atomic Hund's coupling, here the Hund's coupling has a very minimal effect on the competition 
between the inter-moment spin exchange and Kondo coupling. Specifically, in the regime where the inter-$d$-moment exchange interaction dominates,
we can consider the inter-moment spin singlet as being
formed from the Hund's coupled effective moment.
The latter is primarily made up of the $d$-spins 
due to the large $c$-electron bandwidth.
In the Kondo-dominating regime, there is one additional feature that further reduces the effect of 
the Hund's coupling: The leading Kondo coupling is off-site, in contrast to the onsite nature of the Hund's coupling. Thus, the Hund's coupling likewise will not disturb the formation of the Kondo singlet between the $d$- and $c$-spins.

\subsection*{F. Large-$N$ limit} 
\label{sec:largen}
We solve the Kondo-Heisenberg model with the Hund's coupling in a large-$N$ limit. We generalize $SU(2)$ spin symmetry to $SU(N)$ spin symmetry and let $J^H_{\RR,\RR'} \rightarrow  J^H_{\RR,\RR'}/N$ and $J^K_{\RR,\RR_1a_1,\RR_2a_2} \rightarrow J^K_{\RR,\RR_1a_1,\RR_2a_2} /N, J_a\rightarrow J_a/N$. We then introduce Abrikosov fermions ${f}_{\RR,\sigma}$, which satisfy a local constraint $\sum_{\sigma}{f}_{\RR,\sigma}^\dag {f}_{\RR,\sigma}=1$, and rewrite the spin operators as $\bm{S}_{\RR} = \frac{1}{2}\sum_{\sigma,\sigma'}{f}_{\RR,\sigma}^\dag \bm{\sigma}_{\sigma,\sigma'} {f}_{\RR,\sigma'}$. 

We next perform a Hubbard-Stratonovich transformation 
\ba 
H_H 
&\rightarrow &\sum_{\RR,\RR'} \frac{J^H_{\RR,\RR'}}{2}\bigg[ N\chi_{\RR',\RR} \chi_{\RR',\RR} - \sum_\sigma \bigg({f}_{\RR,\sigma}^\dag {f}_{\RR',\sigma} \chi_{\RR',\RR}+\text{h.c.}\bigg) \bigg]  \nonumber \\
H_{J_K} &\rightarrow &\sum_{\RR,\RR_1,\RR_2,a_1,a_2}
J^K_{\RR,\RR_1a_1,\RR_2a_2}
\bigg[N\zeta_{\RR,\RR_2a_2}\zeta_{\RR,\RR_1a_1}^* -\sum_\sigma \bigg( \zeta^*_{\RR,\RR_1a_1}{f}_{\RR,\sigma}^\dag c_{\RR_2,a_2,\sigma} +\eta_{\RR,\RR_2a_2} c_{\RR_1,a_1,\sigma}^\dag f_{\RR,\sigma}\bigg) 
\bigg] \nonumber \\ 
H_{Hund} &\rightarrow & 
\sum_{\RR,a} \frac{J_{a}}{2}\bigg[ -N\zeta_{\RR,\RR a}\zeta_{\RR,\RR a}^* + \sum_\sigma \bigg( \zeta_{\RR,\RR a}^* {f}_{\RR,\sigma}^\dag c_{\RR,a,\sigma} +\text{h.c.}\bigg) \bigg] \, ,
\ea 
where $\chi_{\RR,\RR'}, \eta_{\RR,\RR_1a_1}$ are the bosonic fields used in the decoupling procedure.
The large $N$ limit
leads to saddle point equations
\ba 
&&\chi_{\RR,\RR'}=\frac{1}{N}\sum_{\sigma}\langle {f}_{\RR,\sigma}^\dag {f}_{\RR',\sigma} \rangle
\, ,\nonumber \\
&&\zeta_{\RR,\RR'a}=\frac{1}{N}\sum_{\sigma}\langle {f}_{\RR,\sigma}^\dag c_{\RR',a,\sigma} \rangle \, . 
\label{eq:sc_field}
\ea 
The local constraints $\sum_{\sigma}{f}^\dag_{\RR,\sigma}{f}_{\RR,\sigma}=1$ are satisfied on average by introducing a Lagrangian multiplier $\lambda$: $\lambda (\sum_{\sigma}{f}^\dag_{\RR,\sigma}{f}_{\RR,\sigma}-1)$.
At the saddle point, 
$\lambda$ is determined by requiring 
\ba 
\sum_{\sigma}\langle {f}_{\RR,\sigma}^\dag {f}_{\RR,\sigma} \rangle =1
\label{eq:sc_lam}
\ea 

We solve Eqs.~\ref{eq:sc_field} and ~\ref{eq:sc_lam} self-consistently and derive the phase diagram at zero temperature as shown in the main text.
In the small Fermi-surface phase,
the Luttinger theorem of the Kondo lattice 
is seen as satisfied by
the separate responses of the local moments and conduction electrons \cite{senthil2004a,Piv2004}
to the adiabatic insertion of an external flux \cite{Osh2000}.

\subsection*{G. Symmetry properties in the heavy fermion phase}
\label{sec:higgs}
In the heavy Fermi liquid phase, 
$ \hat{\varsigma}_{\RR,\RR',a}$ condensates.
The condensation locks the phase factors of the pseudo-fermion under the $U(1)$ gauge and $M_x$ mirror 
transformations to those of the physical $c$-electrons.
Thus, $f_{\RR,\sigma}$ acquires not only
a $U(1)$ physical charge
as in the standard Kondo model,
but also 
the $+1$ mirror eigenvalue. Thus the heavy bands induced by the Kondo effect has $+1$ mirror eigenvalue along $\Gamma$-$K$ lines. Consequently, its crossing with wide bands of $-1$ mirror eigenvalue is symmetry-protected. 

\clearpage  
\newpage

\clearpage  
\newpage

\begin{table} 
\centering
\begin{tabular}{c|c | c | c }
\hline 
    Momentum &  $\Gamma$ & $M'$ & $K$\\
    \hline 
    $M_x$ eigenvalue of band 1   & +1&-1 &+1 \\
    $M_x$ eigenvalue of band 2   & +1 &+1 &-1 \\
    $M_x$ eigenvalue of band 3   & -1& +1 & +1  \\
\hline 
\end{tabular}
\caption{$M_x$ symmetry eigenvalues of the top three bands. We label the bands from the top to the bottom as 1,2,3,4,5.}
\label{tab:sym}
\end{table}

\begin{table}[]
    \centering
    \begin{tabular}{ c|c |c | c| c | c| c | c| c }
    \hline 
      $(i,j)$ &  $(0,1)$ & $(-1,0)$ & $(1,0)$ & $(0,-1)$&  $\pm(-2,2)$
      & $(-2,0)$ &$(0,2)$ 
      \\
      \hline 
    $t_{i{a}_1 +ja_2 }^d/t$  & -0.023& -0.023 & -0.023& -0.023 & 0.025 
    & 0.019 & 0.019 
    \\
    \hline 
   &$(2,0)$ &$(0,-2)$ & $(-1,2)$ & $(1,-2)$  & $(1,-2)$ & $(-1,2)$ \\
    \hline 
    &0.019 & 0.019 & -0.017 &-0.017 &-0.017 &-0.017 \\
    \hline 
    \hline 
    $(i,j)$  &$ (1,0) $&$ (0,-1) $& $(1,1)$ & $(-1,-1)$&$ (-1,2) $&$ (-2,1) $& $(1,-2)$ &$ (2,-1)$  \\ 
    \hline 
    $V_{i{a}_1 +ja_2,1 }/t$ &0.066 &-0.066 & -0.025 & 0.025 & 0.047 & -0.047 & 0.040 & -0.040
    \\ 
    \hline 
      &$(2,0)$ & $(0-2)$ &$(-3,2)$ & $(-2,3)$ &$(-3,1)$ & $(-1,3)$ \\ 
    \hline 
   & 0.032 & -0.032 & -0.028 & 0.028 &0.020 & -0.020 
    \\ 
    \hline \hline 
    $(i,j)$ &(0,0) & $(0,1)$ & $(-1,0)$ & $(1,0)$ & $(0,-1)$ &$(1,-1)$ & $(-1,1)$   \\
    \hline 
    $V_{i{a}_1 +ja_2,2 }/t$&-0.04  & 0.0481 & 0.0481 & 0.0457 & 0.0457 &  -0.0758 & -0.0726  \\
    \hline 
     &$(1,1)$ & $(-1,-1)$ & $(2,-2)$ & $(-1,2)$ & $(-2,1)$ & $(2,0)$ & $(0,-2)$ & $(-2,2)$ \\
    \hline 
  & -0.0377 & -0.0377 & 0.0384 & 0.0365 & 0.0365 & 0.0275 & 0.0275 & -0.0242 \\
    \hline \hline 
    $(i,j)$& $(0,1)$ & $(0,-1)$ & $(-1,0)$ & $(1,0)$ & $(-1,1)$ & $(1,-1)$ & $(-1,-1)$ & $(1,1)$ \\
    \hline 
    $t_{ia_1+ja_2,11}^c$ & -0.1768 & -0.1768 & -0.1768 & -0.1768 & -0.0789 & -0.0789 & -0.0378 & -0.0378 \\
    \hline 
    & $(2,-2)$ & $(-2,2)$ \\
    \hline 
    & -0.0403 & -0.0403 \\
    \hline \hline 
    $(i,j)$ & $(0,-1)$ & $(0,1)$ & $(-1,0)$ & $(1,0)$ & $(-1,-1)$ & $(1,1)$ & $(2,0)$ & $(-2,0)$ \\ 
    \hline 
    $t_{ia_1+ja_2,22}^c$ & 0.1389 & 0.1389 & 0.1389 & 0.1389 & 0.0292 & 0.0292 & -0.0383 & -0.0383  \\
    \hline 
    & $(0,-2)$ & $(0,2)$ \\
    \hline 
    & -0.0383 & -0.0383\\ 
    \hline \hline 
    $(i,j)$ &  $(0,0)$ & $(0,-1)$ & $(0,1)$ & $(-1,0)$ & $(1,0)$ & $(-1,-1)$ & $(1,1)$  \\
    \hline 
    $t_{ia_1+ja_2,12}^c$ &2.3657 & 0.1389 & 0.1389 & 0.1389 & 0.1389 & 0.0292 & 0.0292  \\
    \hline 
   &$(2,0)$ & $(-2,0)$  & $(0,-2)$ & $(0,2)$ \\ 
    \hline 
   & -0.0383 & -0.0383 & -0.0383 & -0.0383 \\
    \hline \hline 
    &$E_d/t$ & $E_1/t$ & $E_2/t$ \\
    \hline 
    &0.01 & 0.92 & 1.37 \\ 
    \hline 
     \end{tabular}
    \caption{Parameters of the kinetic term in the Wannier basis. }
    \label{tab:par1} 
\end{table}

\end{document}